\documentclass[journal]{IEEEtran}
\usepackage{graphicx}
\usepackage{amsmath}
\ifCLASSOPTIONcompsoc
\usepackage[caption=false,font=normalsize,labelfont=sf,textfont=sf]{subfig}
\else
\usepackage[caption=false,font=footnotesize]{subfig}
\fi
\usepackage{stfloats}
\begin{document}
\title{\huge PET Quantification of Ultra Low Activity via Inhomogeneous Poisson Process Parameters Estimation Directly from Listmode Data}
\author{Zhenzhou Deng, Xin Zhao, Anyi Li

\thanks{Zhenzhou Deng, Xin Zhao are all with Department of Electronics, College of Information Engineering, Nanchang University, Nanchang, China. Anyi Li comes from Jiluan Academy. Email: dengzhenzhou@gmail.com}
}

\maketitle
\pagestyle{empty}
\thispagestyle{empty}

\begin{abstract}
Metabolic imaging with PET/CT using $^{18}$F-Fludeoxyglucose ($^{18}$F-FDG) as well as other imaging biomarkers has achieved wide acceptance in oncology, cardiology and neurology not only because of the unique metabolic information generated by this modality, but also because of its ability to quantify biological  processes. However, PET quantification is affected by many technical and physiologic factors, and then recognized as an important problem for diagnosis, determination of prognosis, and response monitoring in oncology. In this work, we investigated the effect of reduced PET emission count statistics on the accuracy and precision of tracer quantification, and proposed Inhomogeneous Poisson Process Parameter Estimation (I3PE) method. In I3PE method, we modelled the coincidence event as Inhomogeneous Poisson Process, and estimate its parameter directly from the streaming listmode data. To evaluate the effectiveness, a experiment using $^{18}$F-FDG was implemented on LIGHTNING. The results not only demonstrate I3PE method, but also evaluated the minimal detectable activity of the using PET machine. According $0.3\%$ mean error rate criterion, LIGHTNING can detect several nano-Curie, cooperated with I3PE method.
\end{abstract}

\IEEEpeerreviewmaketitle
\section{Introduction}
Metabolic imaging with PET using $^{18}$F-Fludeoxyglucose ($^{18}$F-FDG) as well as other imaging biomarkers has achieved wide acceptance in oncology,cardiology and neurology because of not only the unique metabolic information generated by the PET modality, but also its ability to quantify biological processes. However, PET quantification is affected by many technical and physiologic factors, and then recognized as an important issue for diagnosis, determination of prognosis, and response monitoring in oncology.\cite{parkash2004potential}. In particular, when the injected tracer activity is reduced to  nano-Curies level, the coincidence detection generated by the injected tracer is severely corrupted by random coincidences.\cite{QP}. To quantify such low activity tracer, there is a substantial interest in moving PET quantification in the direction of analyzing the counting statistics at the earliest possible stage to avoid information-loss during filtering or iteration in the reconstruction process.\cite{PMID}.

In this work, we investigated the effect of reduced PET emission count statistics on the accuracy of tracer quantification, and proposed Inhomogeneous Poisson Process Parameters Estimation (I3PE) method to quantify the ultra low activity directly from listmode data. In this method, the event counts in each LOR (Line Of Response) are modeled as an Inhomogeneous Poisson Process (IPP) whose rate function can be represented using a biased exponent function with time constant of $6588/\ln2 $ s. An estimate of these rate functions is obtained by maximizing the likelihood of the arrival times of each detected photon pair. An experiment with low activity tracer was conducted as follows, and the results demonstrated the effectiveness of I3PE for quantification of ultra low activity.
\section{Method}
\subsection{The Counting Statistics Model}
 The positron emissions from the radioactive source are modeled as IPP. The counting process observed at the detectors is corrupted by randoms and scatters that can also be modeled as IPP. In particular, scintillation detectors employing LYSO have evidential backgrounds, which produce severe random events and limit the quantification of ultra low activity. When the radioactive source is approximately a point source, the affect of scatters deduced by the radioactive source is negligible comparing with randoms component. Combining the random and true components, we have the coincidence count rate model in low activity condition:
\begin{equation}
\label{IPP}
\lambda'_i(t) = \lambda^t_i(t) + \lambda^r_i(t).
\end{equation}
where $\lambda^t_i(t) = \lambda_0 w_i\exp\{-t (\ln2) / 6588\}$ and $\lambda^r_i(t) = C$ are the trues and scatters rate functions for the $i$th detector pair and $\lambda'_i(t)$ is the rate function for the process actually observed at the $i$th detector pair. In the estimation of the rate function parameters, the response weigh $w_i$ for the given radioactivity have been determined through a calibration procedure in normal counting rate and can be treated as known parameters.
\begin{figure}[ht]
\hspace*{\fill}\subfloat{\includegraphics[width=1.4in]{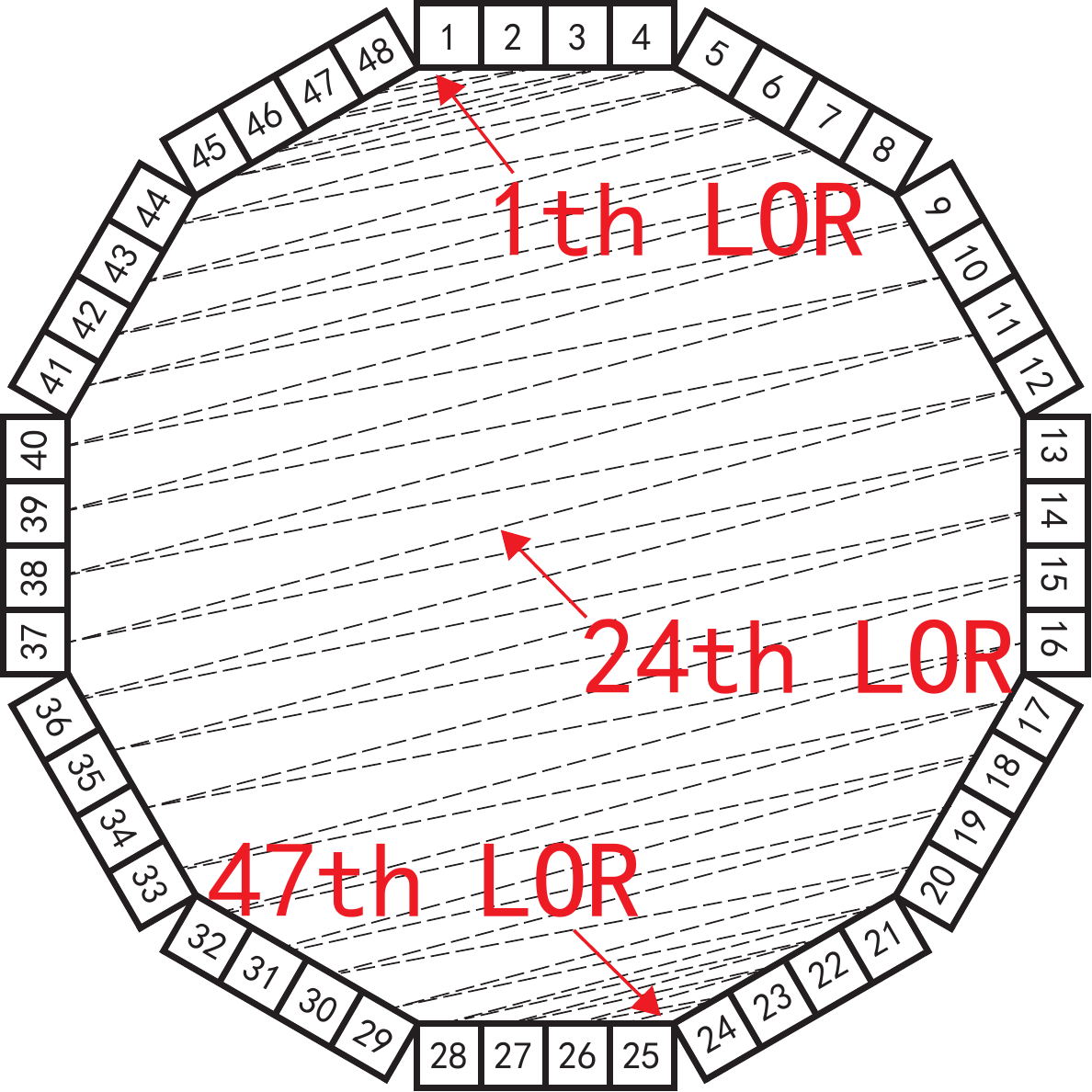}}\hspace*{\fill}\subfloat{\includegraphics[width=1.4in]{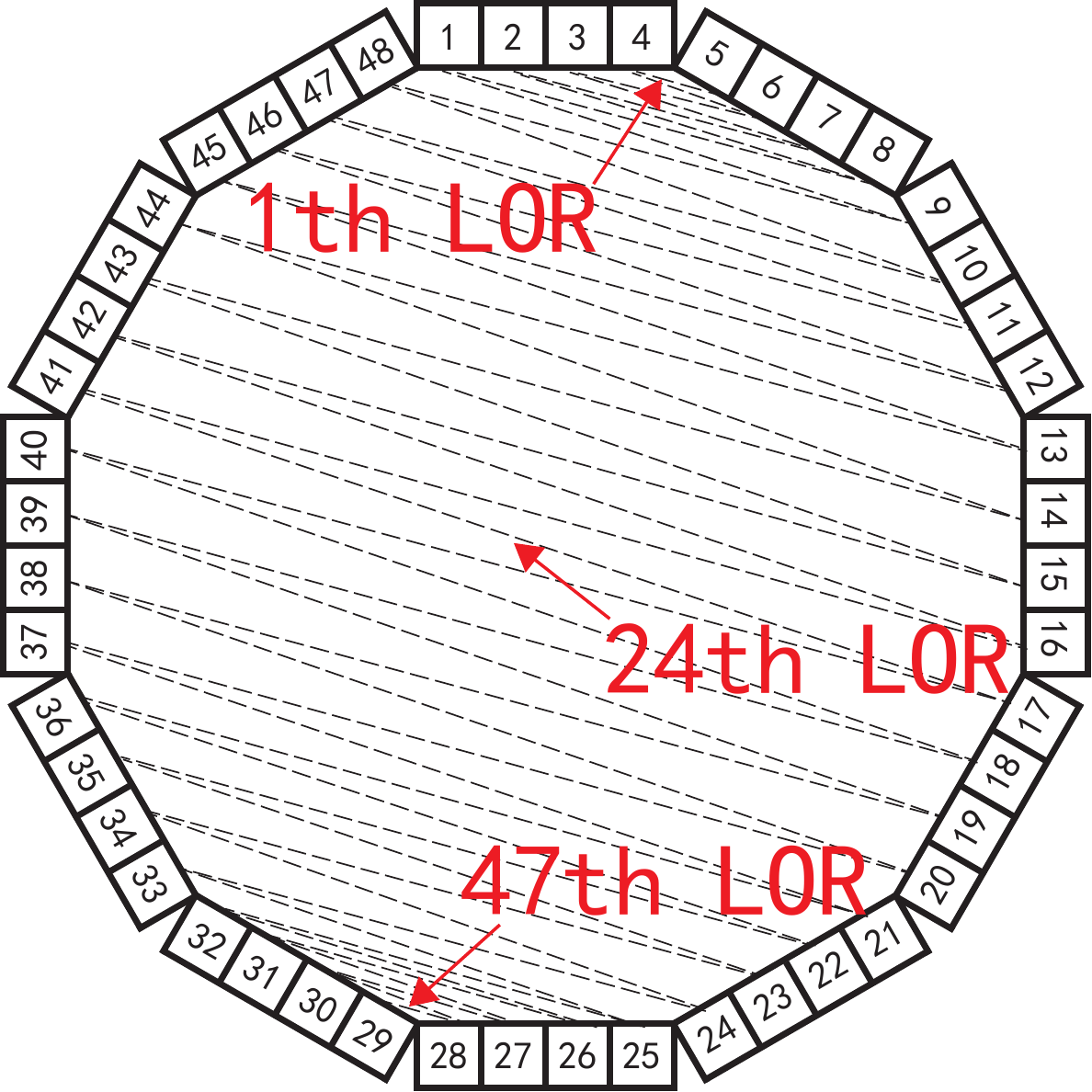}}\hspace*{\fill}
\caption{The arrangement of the coincidence listmode data. (Left) LORs of 0-degree View. (Right) LORs of 30-degree View.}
\label{view}
\end{figure}
\subsection{Inhomogeneous Poisson Process Parameters Estimation Method}
We estimate the required parameters $\lambda_0$ and $C$ in Eq. (\ref{IPP}) according to Max Likelihood criterion from the obtained coincidence listmode $\{t_{i,j}\}$. $\{t_{i,j}\}$ are the streaming data, and $j$ is the index of LORs, which are arranged as Fig. \ref{view}. The parameters estimation was conducted by Majorize-Minimization iteration alternately for $\lambda_0$ and $C$ as follow:
\begin{equation}
\label{Iter1}
\lambda_{0,k+1} = arg\  max \ \ L(\{t_{i,j}\};C_k),
\end{equation}
\begin{equation}
\label{Iter2}
C_{k+1} = arg\  max \ \ L(\{t_{i,j}\};\lambda_{0,k+1}),
\end{equation}
where the $L(\cdot)$, $C_k$ and $\lambda_{0,k}$ are the likelihood function, the calculated $C$ and $\lambda_{0}$ of the $k$th iteration, respectively.

\subsection{Bayesian Statistics Parameters Estimation Method}
On the basis of $\lambda_{0,k+1}$ and $C_{k+1}$, We take the Bayesian Statistics Method based on Max Likelihood criterion to estimate $\lambda_0$ and $C$ from the listmode $\{t_{i,j}\}$ again. Parameter estimation is for $ \lambda_{0, k + 1} $ and $ C_ {k + 1} $ carried out as follows:
\begin{equation}
\label{Iter3}
\hat{\lambda_0} = arg\ max \ \ B(\{t_{i,j}\};\lambda_{0,k+1}),
\end{equation}
\begin{equation}
\label{Iter4}
\hat{C} = arg\  max \ \ B(\{t_{i,j}\};C_{k+1}),
\end{equation}
where the $B(\cdot) = L(\cdot)P_p(\cdot)$, $P_p(\cdot)$ is prior probability, $\hat{C}$ and $\hat{\lambda_{0}}$ are estimation acquired from the bayesian statistics.
\begin{figure}[ht]
\centering
\subfloat{\includegraphics[width=1.6 in]{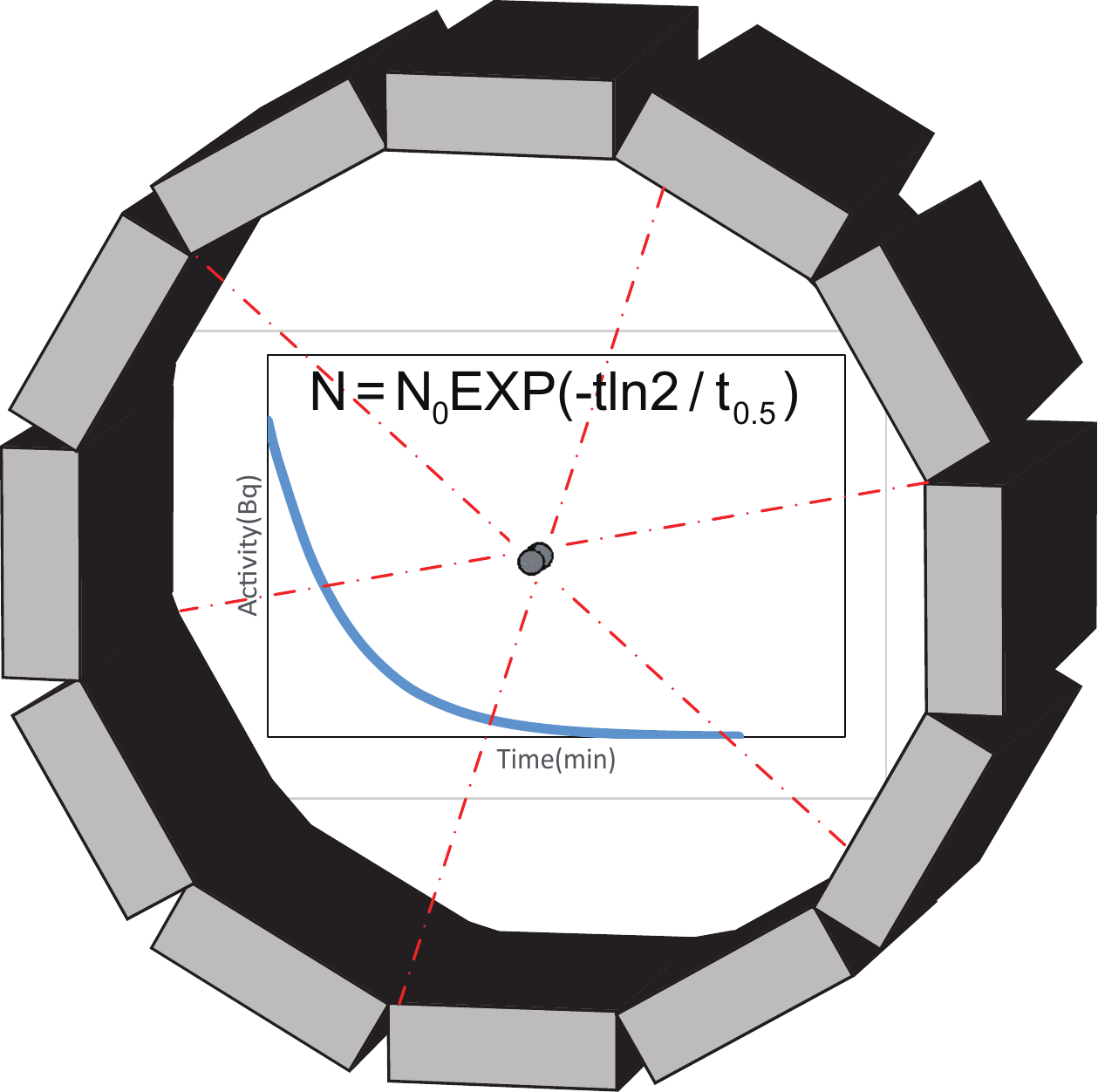}}\hspace*{\fill}
\subfloat{\includegraphics[width=1.7 in]{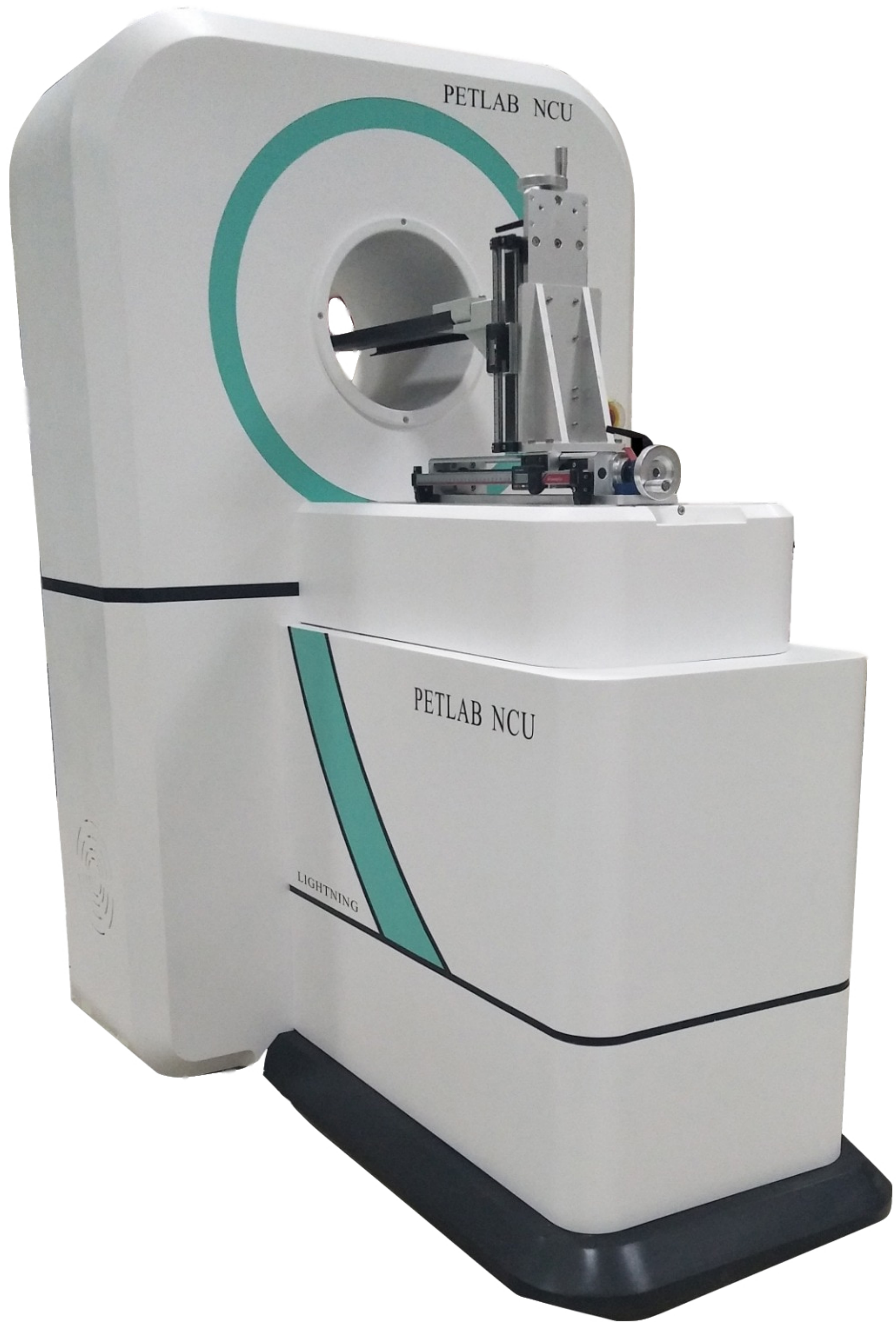}}
\caption{(Left) Scheme of the scanner geometry employed in this work. (Right) The experiment setup using the low activity positron emitter.}
\label{illustrater}
\end{figure}

\section{Experiment and Results}
To evaluate the performance of I3PE method for the low activity positron emitter, a continuous experiment was implemented on a preclinical scanner--LIGHTNING, which mainly consists of twelve independent detector modules. Each detector module of size $13\times26.5\times106$ $mm^{3}$ contains $1\times4$ array of lutetium yttrium oxyorthosilicate (LYSO)/silicon photomultiplier (SiPM) blocks. We configured the activity of the positron emitter by dilution and decay, which kept the accuracy of the given radioactivity. The experiment lasted 16 hr. and 19 min. totally. The procedures were executed as follow:

(i) \emph{Radioactive Source Preparation}      We prepared FDG mother liquor with activity of 92 uCi and volume of 88 ml. The activity was measured by radioactivity meter RM-905a. We collected 0.3 ml liquor from the mother liquor, obtaining 11605 Bq FDG.

(ii)  \emph{Radioactive Source Positioning}     The prepared source was positioned paralleled to the axial direction and located at the center of the FOV. The position was indicated with laser positioning as in Fig. \ref{illustrater}.

(iii) \emph{Data Collection with Radioactive Decay}  We executed the PET scanning along with the radioactivity decay. The activity of the first data collection was 10093 Bq, which is owing to the $^{18}$F decay of 1329 s. During the scanning, the raw event data were all transmitted to and restored in the computer workstation. The encoding of the LORs of the listmodes is shown as Fig. \ref{view}.

\begin{figure}[ht]
\centering
\subfloat{\includegraphics[width=1.3in]{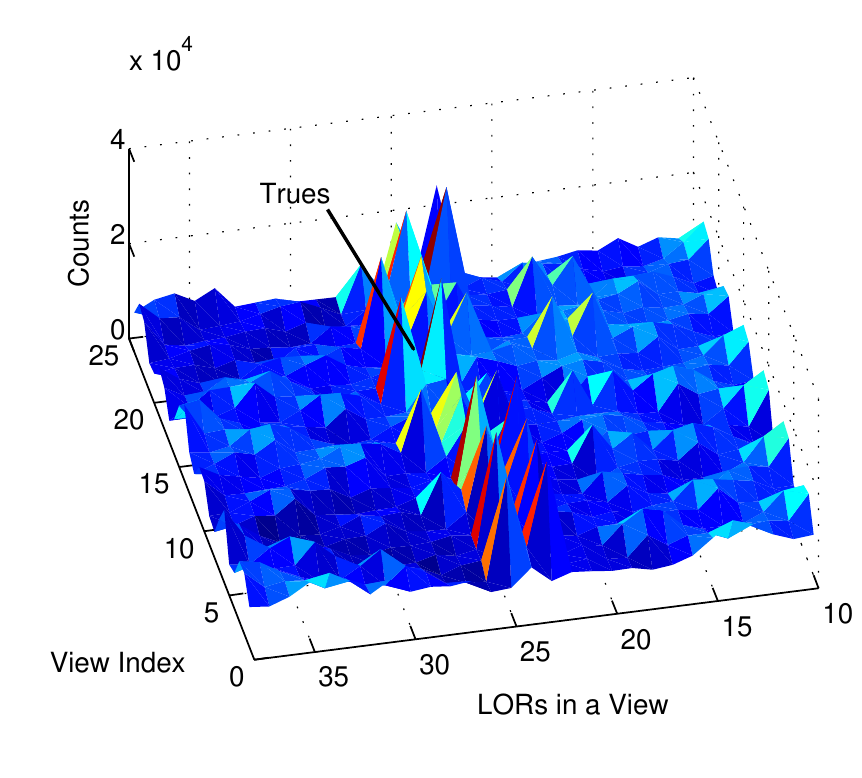}}\hspace*{\fill}
\subfloat{\includegraphics[width=1.3in]{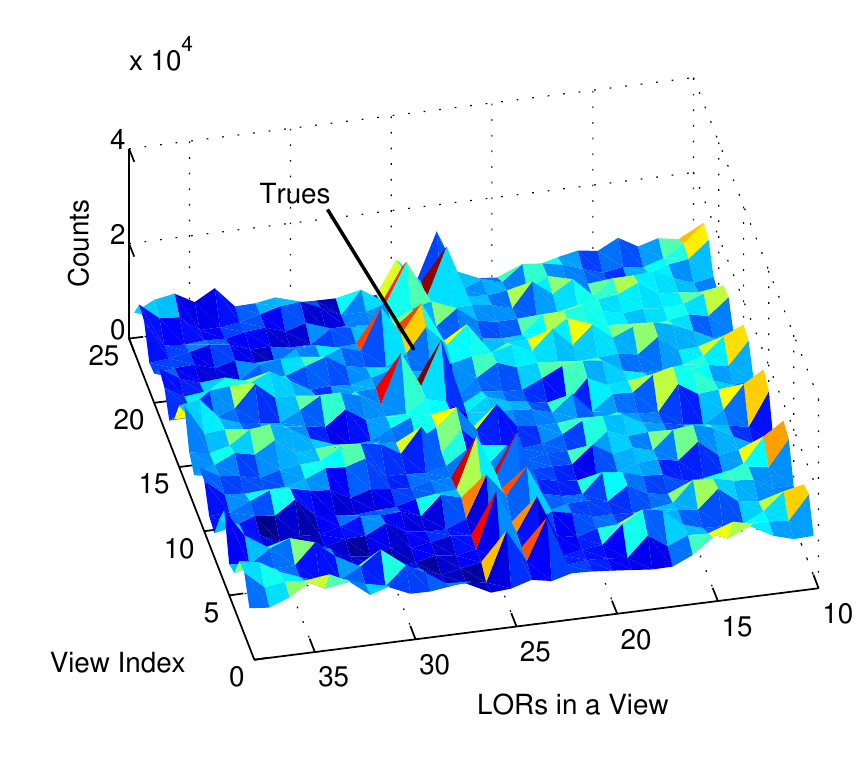}}\hspace*{\fill}
\subfloat{\includegraphics[width=1.3in]{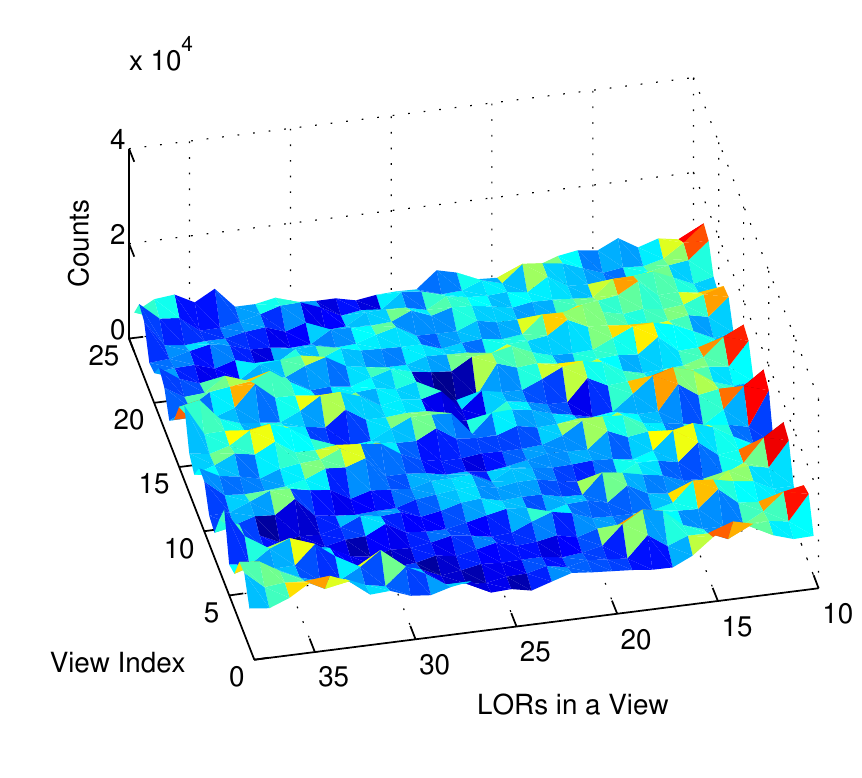}}
\caption{The histogram of the raw listmode data. (Left) 100 nCi activity. (Middle) 50 nCi activity. (Right) 5 nCi activity.}
\label{visibleFLOOD1}
\end{figure}

In data analysis, whether the positron emitters can be detected by PET system is derived by statistical analysis. Fig. \ref{visibleFLOOD1} shows the histogram of the raw listmode data. Table \ref{Minimal_Detectable_Activity} shows the minimal detectable activity with different scanning durations and Mean Error Rates (MER).
\begin{table}[h]
\caption{The Minimal Detectable Activity }
\label{Minimal_Detectable_Activity}
\centering
\begin{tabular}{c c c}
\hline
Scanning Duration (min) & MER(\%) & Minimal Detectable Activity (nCi)\\
\hline
5  & 0.3 &  125.3 \\
10 & 0.3 &  98.3 \\
50 & 0.3 &  59.8  \\
5  & 5 &  25.3 \\
10 & 5 &  18.3 \\
50 & 5 &   7.3 \\
\hline
\end{tabular}
\end{table}

\section{Conclusion and Discussion}
We proposed the I3PE method to quantify the tracer of ultra low activity directly from the listmode data. In this method, we analyzed the counting statistics of ultra low activity and modeled the coincidence events as Inhomogeneous Poisson Process with specified rate function. A Majorize-Minimization iteration was constructed to estimate the parameters of the rate function. The experiment results show that the proposed I3PE method has the capability to perform quantitative measurements either in very short PET scanning duration or with low source activities. Considering the spatial inhomogeneity in PET, more results with enough testing phantoms will be presented at the conference.

\bibliographystyle{IEEEtran}

\bibliography{IEEEabrv,QPMVT}

\begin{thebibliography}{1}
\providecommand{\url}[1]{#1}
\csname url@samestyle\endcsname
\providecommand{\newblock}{\relax}
\providecommand{\bibinfo}[2]{#2}
\providecommand{\BIBentrySTDinterwordspacing}{\spaceskip=0pt\relax}
\providecommand{\BIBentryALTinterwordstretchfactor}{4}
\providecommand{\BIBentryALTinterwordspacing}{\spaceskip=\fontdimen2\font plus
\BIBentryALTinterwordstretchfactor\fontdimen3\font minus
  \fontdimen4\font\relax}
\providecommand{\BIBforeignlanguage}[2]{{%
\expandafter\ifx\csname l@#1\endcsname\relax
\typeout{** WARNING: IEEEtran.bst: No hyphenation pattern has been}%
\typeout{** loaded for the language `#1'. Using the pattern for}%
\typeout{** the default language instead.}%
\else
\language=\csname l@#1\endcsname
\fi
#2}}
\providecommand{\BIBdecl}{\relax}
\BIBdecl

\bibitem{parkash2004potential}
R.~Parkash, T.~Ruddy, A.~Kitsikis, R.~Hart, L.~Beauschene, K.~Williams,
  R.~Davies, M.~Labinaz, R.~Beanlands \emph{et~al.}, ``Potential utility of
  rubidium 82 {PET} quantification in patients with 3-vessel coronary artery
  disease,'' \emph{Journal of nuclear cardiology}, vol.~11, no.~4, pp.
  440--449, 2004.

\bibitem{QP}
D.~Zhenzhou and Q.~Xie, ``Quadratic programming time pickoff method for
  multivoltage threshold digitizer in {PET},'' \emph{IEEE Transactions on
  Nuclear Science}, vol.~62, no.~4, pp. 1--9, 2015.

\bibitem{PMID}
D.~Zhenzhou, Q.~Xie, Z.~Duan, and P.~Xiao, ``Scintillation event energy
  measurement via a pulse model based iterative deconvolution method,''
  \emph{Physics in medicine and biology}, vol.~58, pp. 7815--7827, 10 2013.

\end{thebibliography}
\end{document}